\documentclass[pra,aps,amsfonts,amsmath,superscriptaddress,%
  twocolumn,showpacs,floatfix]{revtex4}
\usepackage{amsfonts}
\usepackage{amsmath}
\usepackage{bm}
\usepackage{epsfig}
\usepackage{mathrsfs}

\usepackage[usenames]{color}
\usepackage{ulem}

\begin{document}

 \title {\bf Second random-phase approximation with the Gogny force. First applications.}
\author{D. Gambacurta}
\address{Grand Acc\'el\'erateur National d'Ions Lourds (GANIL), CEA/DSM-CNRS/IN2P3, Bvd Henri Becquerel, F-14076, Caen, France}
\author{M. Grasso}
\address{Institut de Physique Nucl\'eaire,
 Universit\'e Paris-Sud, IN2P3-CNRS, F-91406 Orsay Cedex, France}
\author{V. De Donno}
\author{G. Co'}
\address{   
Dipartimento di Matematica e Fisica ``E. De Giorgi'',  
Universit\`a del Salento and
INFN,  Sezione di Lecce, 
I-73100 Lecce, Italy}
\author{F. Catara}
\address{Dipartimento di Fisica e Astronomia and INFN, Via Santa 
 Sofia 64, I-95123 Catania, Italy }


\begin{abstract}

We present the first applications of the second
random-phase-approximation model with the finite-range Gogny
interaction. We discuss the advantages of using such an interaction in
this type of calculations where 2 particle-2 hole configurations are
included.  The results found in the present work confirm the well
known general features of the second random-phase approximation
spectra: we find a large shift, several MeV, of the response
centroids to lower energies with respect to the corresponding
random-phase-approximation values. As known, these results indicate
that the effects of the 1 particle-1 hole/2 particle-2 hole and 2
particle-2 hole/2 particle-2 hole couplings are important.  It has
been found that the changes of the strength distributions with respect
to the standard random-phase-approximation results are particularly
large in the present case. This important effect is due to some
large neutron-proton matrix elements of the interaction and indicates
that these matrix elements (which do not contribute in the mean-field
calculations employed in the conventional fit procedures of the force
parameters) should be carefully constrained to perform calculations
beyond the mean-field approach.
 
\end{abstract}

\vskip 0.5cm  

\maketitle


The new and complex phenomenology associated to exotic neutron-rich
and neutron-deficient isotopes is enriching the experimental landscape
in nuclear physics.  In the perspective of improving the theoretical
description of exotic nuclei, efforts are concentrated on the
refinement of the existing many-body theories, as well as on the
microscopic derivation of the nuclear interaction.  Both aspects are
fundamental and complementary to improve our description of the
nucleus as a many-body interacting system.  In the context of the
Energy Density Functional theories, complex configurations and
correlations are included nowadays within several theoretical
frameworks beyond the conventional mean-field models based on both the
Skyrme and the Gogny effective interactions as well as on relativistic
formulations (e.g. in Refs.
\cite{bender,colo,pillet,libert,litvinova,egido}).

Within the scheme of the standard random-phase approximation (RPA) the
excited states are described as superpositions of 1 particle-1 hole
($1ph$ ) and 1 hole-1 particle ($1hp$) configurations.  Evidently,
those excited states containing non negligible multiparticle-multihole
components are not well described by this theory.  Furthermore, the
width of the excited states cannot be reproduced except for the
single-particle Landau damping and for the escape width (if continuum states are taken into 
account).  A well known extension of the RPA
scheme is the second RPA (SRPA) model which is obtained with the
inclusion of the 2 particle-2 hole ($2ph$) configurations.  This leads
to a richer treatment of the excitation modes.  The spreading width
can also be described owing to the coupling with the $2ph$
configurations.

After having presented in previous works some applications of the SRPA
theory with the Skyrme zero-range interaction
\cite{gamba10,gamba11a,gamba11b} we present here the first
applications of the SRPA theory with the Gogny force.  Although the
use of a finite-range interaction turns out to be numerically more
demanding with respect to the zero-range case, it presents some
advantages.  The first one is related to the fact that the Gogny force
has been introduced and adjusted to be used in both the Hartree-Fock
and the pairing channels.  Since in the SRPA theory not only the
standard RPA-type particle-hole ($ph$) matrix elements of the
interaction are present, it seems to us that the use of a force
tailored to handle also other kinds of terms, such as the
particle-particle matrix elements, is more appropriate.  A second, non
negligible, advantage is the finite range of the four central terms of
the Gogny force.  We expect that this feature provides, in a natural
way, convergent results with respect to the increase of the energy
cutoff in the $2ph$ space of the SRPA calculations.  We do not expect
a full convergence because the Gogny interaction contains also
density-dependent and spin-orbit terms which are of zero-range type.
We remark that our SRPA calculations are not fully self-consistent
because we neglect the Coulomb and the spin-orbit terms in the
residual interaction.  This means that the dependence of the results
on the numerical cutoff is related in this work only to the zero-range
density-dependent part of the interaction.  The cutoff-dependence of
the Skyrme-SRPA results has been addressed and discussed in
Ref. \cite{gamba10}. 

The details of the SRPA formalism may be found in the literature.  The
derivations of the SRPA secular equations carried out by using the
method of the equations of motion \cite{yann} or by considering the
small-amplitude limit of the time-dependent density-matrix theory
\cite{tohyama,lacroix} provide expressions of the matrices $A$ and $B$
which are valid in cases where the interaction does not depend on the
density.  The case of a density-dependent force in extended RPA
theories has been discussed in some early works \cite{war,adachi}. The
specific case of the SRPA model has been considered in
Ref. \cite{gamba11a}, where a prescription to treat the rearrangement
terms of the residual interaction in matrix elements beyond RPA has
been derived. This prescription has been applied also in the present
work. In Ref. \cite{yann} it is demonstrated that the energy-weighted
sum rules (EWSR's) are satisfied in the SRPA model. As a consequence
the first moment obtained in  RPA and SRPA are the same.

To present the first applications where the Gogny interaction is
employed in SRPA calculations we have chosen the nucleus
$^{16}$O. This allows us to better control the heavy numerical problem
associated with the diagonalization of the SRPA matrix. Furthermore,
the study of this nucleus allows us a direct comparison of our SRPA
results with previous Skyrme-SRPA results \cite{gamba10}. In our
calculations we used the D1S parametrization of the Gogny interaction
\cite{D1S,D1S2}.

As an illustration, in the present article we show only results
regarding the monopole isoscalar response.  The calculations are
performed in spherical symmetry in the harmonic oscillator basis. 
The single-particle
and the $1ph$ spaces have been chosen large enough to ensure that the
values of the EWSR's are stable.  All the single-particle states with
an unperturbed energy lower than 60 MeV (that is, all the $1ph$
configurations with an unperturbed excitation energy up to 100 MeV)
are included in the calculations.  As
anticipated, the Coulomb and the spin-orbit contributions are not
taken into account in the residual interaction.   
This leads to a violation of 5 \% of the EWSR's 
within the RPA scheme.  
In the $2ph$ space, we have
considered all the configurations with an unperturbed energy lower
than an energy cutoff $E_{cut}$ and we have studied the numerical
stability of the results with respect to the choice of the cutoff.
In the SRPA calculations with $E_{cut}$ = 60 MeV  
a full check can be done numerically and we have  
found a violation of the EWSR's of 5\% (the same as in the RPA case). 
This violation, as in RPA, is related to the fact that the Coulomb and spin-orbit contributions 
are neglected. 
We have also checked in this case that as expected
the RPA and SRPA first moments are the same. In the other SRPA calculations
where $E_{cut}$ is larger, 
a numerical check of the EWSR's up to an excitation energy equal to $E_{cut}$
cannot be performed because the calculations become numerically extremely heavy.
This is why in the calculations where $E_{cut}$ is larger than 60 MeV, only the low part of the spectrum,
 up to an energy of 40 MeV, is calculated.  
We have verified for $E_{cut}$ = 80 MeV that 85 \% of the EWSR's is found up
to the excitation energy of 40 MeV the rest being located at higher energy.

\begin{figure}[htbp]
\begin{center}
\vspace{0.9cm}
\includegraphics[scale=0.4]{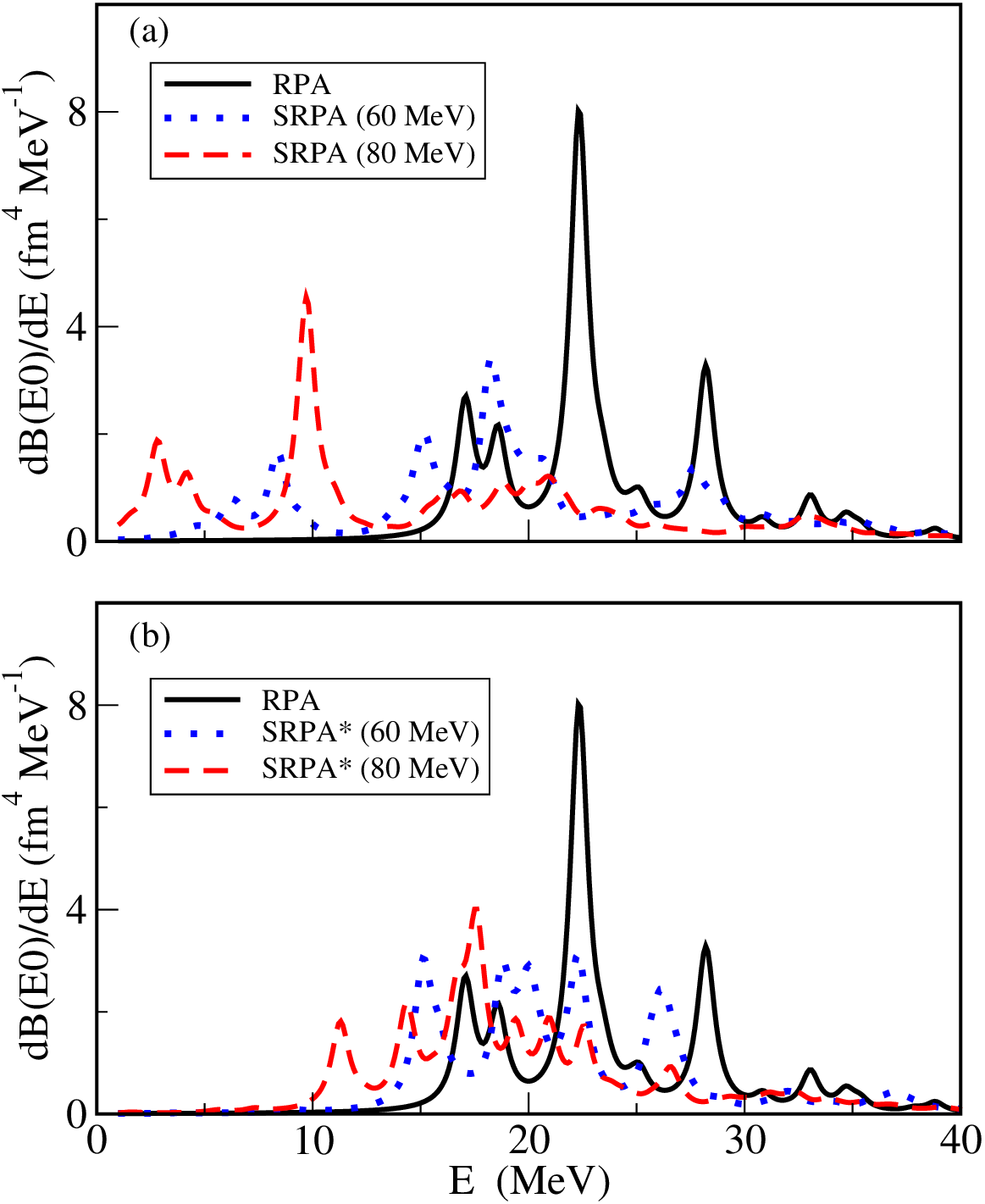} 
\vspace{0.5 cm}
\caption{(Color online) (a) Isoscalar monopole response for the
  nucleus $^{16}$O calculated with the Gogny-RPA model (full line)  
  and with the SRPA approach with an energy cutoff on the $2ph$
  configurations of 60 (dotted line)
  and 80 (dashed line) MeV.
(b) Same as in (a) but in the     
SRPA* scheme. See the text for more details.
}
\label{fig1}
\end{center}
\end{figure}

In these first applications of the Gogny-SRPA model we have found that
some neutron-proton ($\nu\pi$) matrix elements of the interaction,
appearing in the beyond-RPA block matrices, are very large, some of
them being from 5 to 10 times larger than all the other matrix
elements. These matrix elements, that are absent in the standard RPA
calculations, have a strong impact on the stability of the results, in
particular on the peak structure of the response. As we shall show
below, their effects are especially strong in the matrix elements
coupling $1ph$ and $2ph$ configurations. To analyze and single out these 
effects we have performed two different kinds of calculations: (a) a
full SRPA calculation where all the $2ph$ configurations are included;
(b) a calculation performed by considering only the $2ph$
configurations that are composed by pure neutron or proton
excitations. This means that in the case (b) we do not include the
$2ph$ configurations where the two particles and the two
holes have a different isospin nature.  As a consequence, no $\nu\pi$ matrix elements of the
interaction are present in the SRPA matrices in the case (b). 
The $\nu\pi$ matrix elements would appear (i) in the case where the  $2ph$
configurations were  composed by 1 pure neutron and  1 pure proton $1ph$
configurations (standard RPA $1ph$ configurations); (ii)  in the case where both $1ph$ configurations 
were  $\nu\pi$ $1ph$
configurations (typical charge-exchange $1ph$ configurations). 
In the case of the $A_{1ph,2ph}$ matrix we have checked that the majority
of the matrix elements are relatively small (of the order of 0.2-0.7 MeV),
the mean value being  0.2 MeV.
However, some  $A_{1ph,2ph}$ matrix elements (around 100 when $E_{cut}=80 MeV$) are much larger (up to 10 times)
and, in particular, the largest ones are due to the presence of 3 hole-1 particle $\nu\pi$ matrix elements of the
residual interaction of the type: 

\begin{equation}
\label{v1}
 \langle \nu^{-1} \pi |V| \nu^{-1} \pi^{-1} \rangle_A
\end{equation} 

\begin{equation}
\label{v2}
 \langle \pi^{-1} \nu |V| \nu^{-1} \pi^{-1} \rangle_A
\end{equation}
where 'A' stands for 'Antisymmetrized'.
The  angular momentum coupling is done between the first-third and
second-fourth indices. In the largest $A_{1ph,2ph}$ terms 
 the strongest contributions are matrix elements of the residual
interaction of the type  (\ref{v2}) (charge-exchange type). We stress that also some 3
hole-1 particle matrix elements of the residual interaction involving
only neutron or proton states  are larger than the typical ones. However, 
 they are few and we checked numerically that the strong changes in the SRPA response are
not related to them. 
In the approximation (b), also the $\nu\pi$ matrix elements of the matrix
$A_{2ph,2ph}$ (the matrix that couples among themselves the $2ph$
configurations) are neglected. However, these matrix elements are not expected
to have a strong impact and this will be shown later in this work. In the
following we will indicate the two calculations (a) and (b) as SRPA
and SRPA$^*$, respectively. The strong impact of the $\nu\pi$ matrix
elements of the interaction can be seen in Fig. \ref{fig1} where the
isoscalar monopole response for the operator
\begin{equation}
 F_{}^{IS}= \sum  r_i^2 Y_{0 0}(\hat{r}_i),
\label{isos-oper}
\end{equation}
calculated in the SRPA (a) and in the SRPA* 
(b) scheme is displayed for two values of the cutoff 
energy, 
$E_{cut}=$ 60 and 80 MeV. The corresponding Gogny-RPA results are also plotted in the two panels 
of the figure.  
%
To make simpler the comparison between different results, we have
folded the discrete spectra with a Lorentzian function with a width of
1 MeV.  We see that in the SRPA scheme the responses associated with
the different cutoff values are appreciably different and, for
$E_{cut} = 80$ MeV, the main peak of the response is pushed at
energies more than 10 MeV lower than in the RPA case (a). The SRPA*
results of panel (b) are much more stable with respect to the change
of the cutoff energy. This can also be seen by considering the
centroid energies of the strength distributions.  When the energy
cutoff is increased from 60 to 80 MeV the centroid goes from 20.37 to
15.30 MeV (deviation of 25\%) in the full SRPA calculations whereas it
is much less shifted, from 23.97 to 22.37 MeV (7\%), in the SRPA*
case.   It is also worth noticing that in the latter case the
  difference between the spectra corresponding to the two energy cutoff is
  essentially just a shift, while when the $\nu\pi$ matrix elements are
  not neglected, the SRPA strength distribution is very much different
  from the RPA one.  To check more in detail the stability of the
results in the SRPA* case, we have performed also calculations with
cutoff values of 100 and 120 MeV (Fig. 2). When the cutoff is changed
from 80 to 100 MeV the centroid is shifted from 22.37 to 21.32 MeV
(5\%) and when the cutoff is changed from 100 to 120 MeV the centroid
moves from 21.32 to 20.49 MeV (4\%). On the contrary, the SRPA
results still change significantly increasing the energy cutoff and for
values larger than 80 MeV the solution of the corresponding equations
is affected by the presence of a few imaginary eigenvalues. We conclude that the stability
expected when the Gogny interaction is employed seems to be
achieved in the SRPA* case where, by construction, all the large
$\nu\pi$ matrix elements of the residual interaction in the beyond RPA
blocks of the matrices are neglected. 
This indicates that the density-dependent zero-range part of the interaction does not 
affect strognly the convergence of the results. 
The strong impact of these
matrix elements can also be seen in Fig. 3 where we plot the discrete
spectra in a logarithmic scale to emphasize the fragmentation of the
response.  The SRPA and SRPA* spectra (full lines) correspond to a
cutoff of 80 MeV.  In the two panels of this figure, as a comparison,
also the corresponding RPA results are plotted (dashed lines).  We
observe that not only the energies but also the fragmentation of the
peaks is strongly affected in the SRPA response that is shown in
(a). The monopole case is presented here as an illustration. However,
we have verified that also in the dipole and quadrupole cases the
spectra (energies and fragmentation) are strongly modified in the full
SRPA scheme where all the $\nu\pi$ matrix elements of the residual
interaction are included. 

\begin{figure}[htbp]
\vspace{0.9cm}
\includegraphics [scale=0.3] {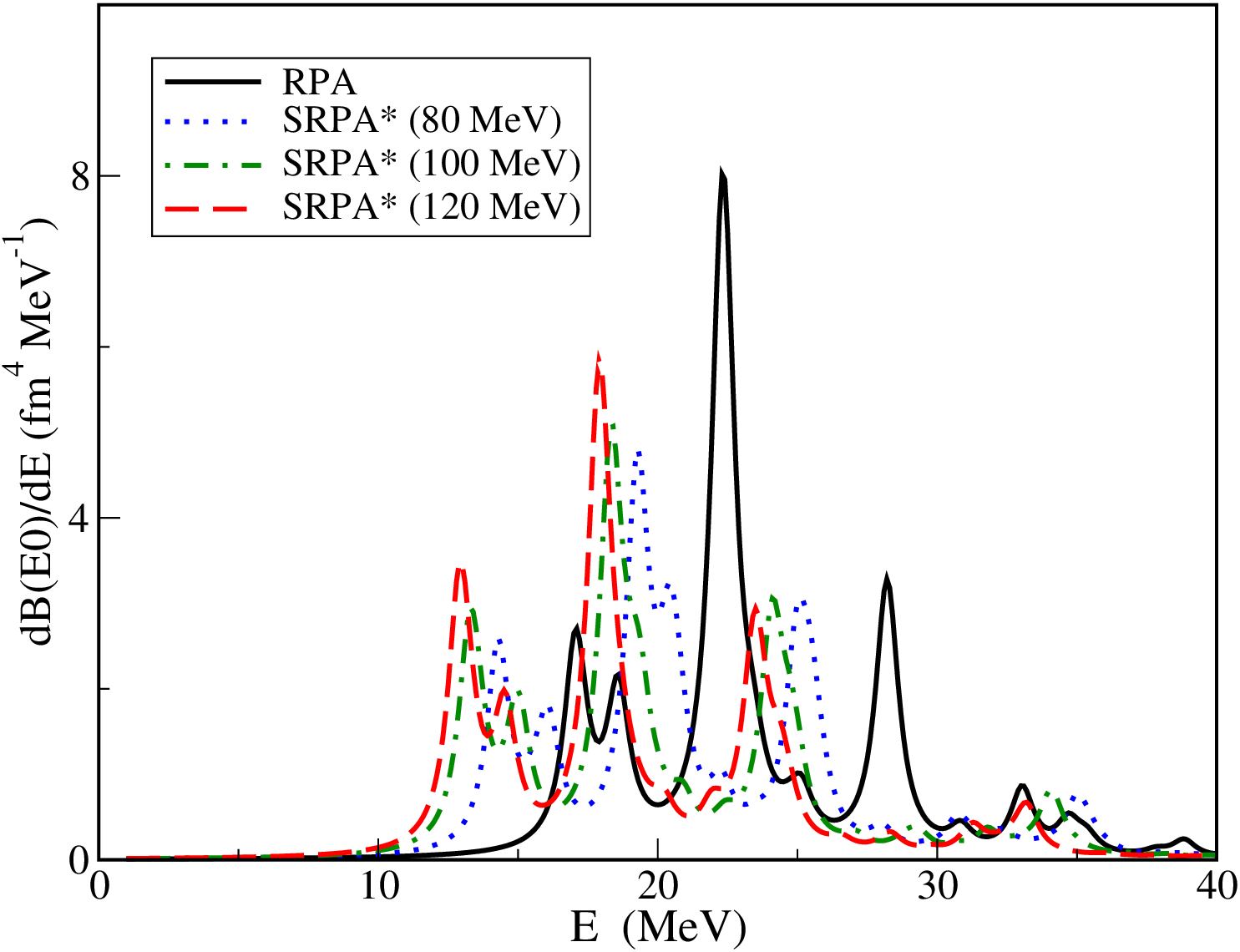}
\vspace{0.9 cm}
\caption{(Color online) 
Isoscalar monopole response for the nucleus $^{16}$O
calculated in the SRPA* case with 
cutoff energies of 80 (dotted line), 100 (dot-dashed line) and 
120 (dashed line) MeV.
The Gogny-RPA results are also plotted (full line).}
\label{fig2}
\end{figure}

\begin{figure}[htbp]
\vspace{0.9cm}
\includegraphics [scale=0.4] {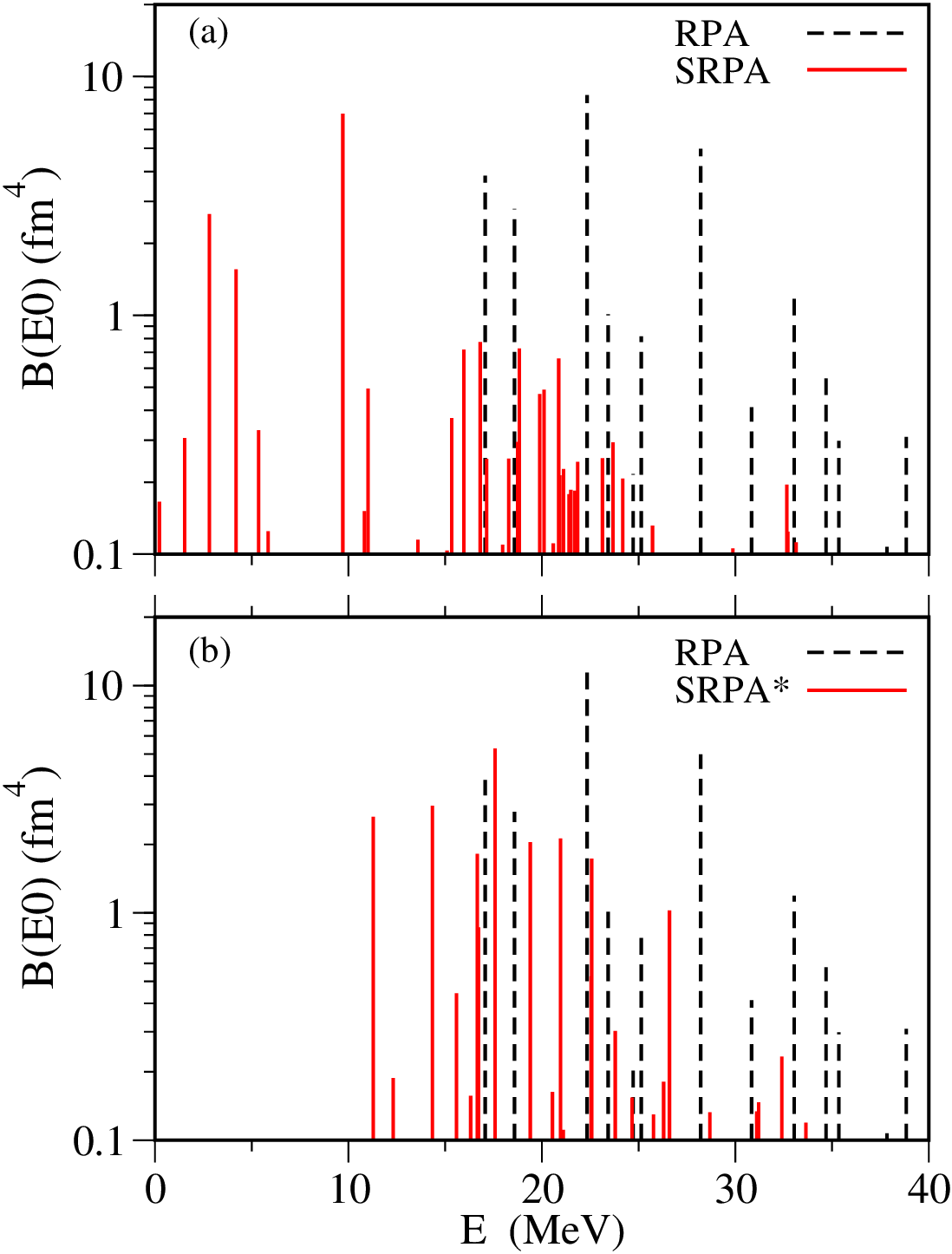} 

\vspace{0.9 cm}
\caption{(Color online) 
Isoscalar monopole discrete strength distributions for the nucleus $^{16}$O
calculated within the SRPA model (full line) with a
cutoff energy of 80 MeV (a). (b) Same as in (a) but in the SRPA* scheme.
{\sl In both panels} , the Gogny-RPA results are also plotted (dashed lines)  
as a reference.}
\label{fig3}
\end{figure}

For the case SRPA*, where a reasonable convergence is achieved with respect to the 
cutoff in the $2ph$ space and the anomalous $\nu\pi$ matrix elements are 
neglected, we can compare with the existing experimental data for the giant monopole 
resonance \cite{expmono}. The experimental centroid is located at 21.13 MeV. 
The centroid energy is equal to 23.88 and 20.49 MeV in RPA and in SRPA*, respectively. 
This indicates that the shift to lower energies obtained in the SRPA* model leads to 
results that are in better agreement with the experimental values.  

The effect we have just pointed out, related to the presence of some
very large $\nu\pi$ matrix elements, is not so surprising.  These
matrix elements do not contribute at the mean-field level where the
fit procedures are commonly performed to select the values of the
parameters of the effective interactions such as the Skyrme and Gogny
forces. For this reason, we cannot exclude that their inclusion in
actual calculations may have an unexpected and strong impact.  It is
also interesting to remark that analogous important effects related to
large $\nu\pi$ matrix elements have been found in the Gogny case also
in recent applications of the variational multiparticle-multihole
configuration interaction theory to the low-lying spectroscopy of the
nucleus $^{30}$Si \cite{pillet2}.  These findings are coherent with
our results and suggest that these matrix elements of the interaction
should be carefully tuned in the fit procedures.

By comparing the present results with the corresponding Skyrme results
of Ref. \cite{gamba10} we observe  that also in that case a
  similar behavior is found, although less pronounced.  This effect
of the SRPA theory is found not only in nuclear physics when
phenomenological interactions like the Skyrme and Gogny forces are
used. This effect is present also in nuclear calculations which employ
forces derived from realistic interactions \cite{papa1,papa2}, as well
as in a completely different domain, that is, in calculations carried
out for metallic clusters (where the interparticle interaction is the
Coulomb force) \cite{gamba09}.  It is important to underline that the
shift found in the Gogny case is comparable to the corresponding
Skyrme result only when all the large $\nu\pi$ matrix elements are
omitted in the Gogny case (SRPA* scheme). Otherwise, their effect is
too strong in pushing the centroid energies to lower values.

\begin{figure}[htbp]
\vspace{0.9cm}
\includegraphics[scale=0.4]{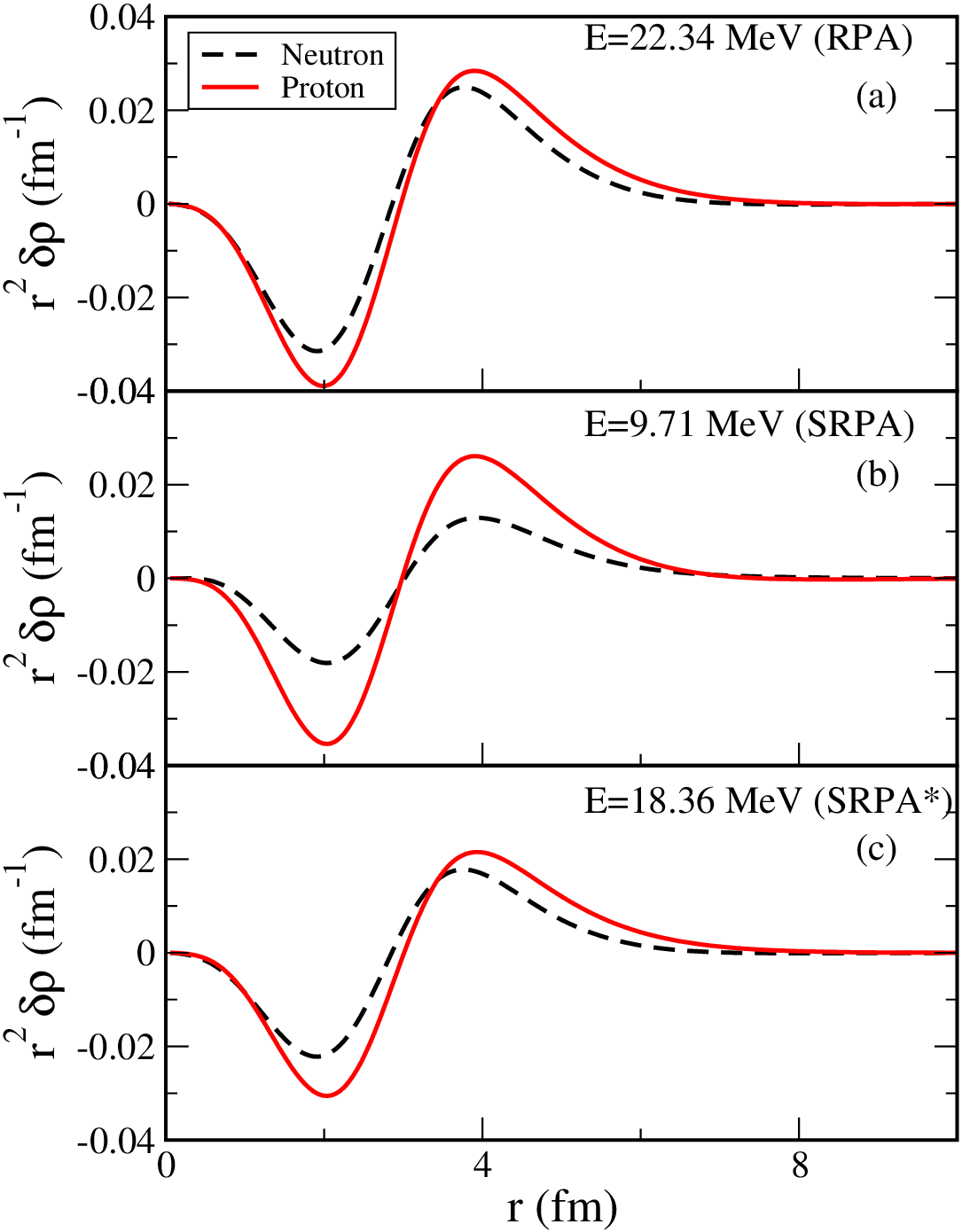} 
\vspace{0.9cm}
\caption{ (Color online)
Neutron (dashed lines) and proton (full lines) transition densities
for the main peaks of the RPA (a), SRPA (b)
and SRPA* (c) strength distributions. 
}
\label{fig4}
\end{figure}
%
\begin{figure}[htbp]
\vspace{0.9cm}
\includegraphics[scale=0.4]{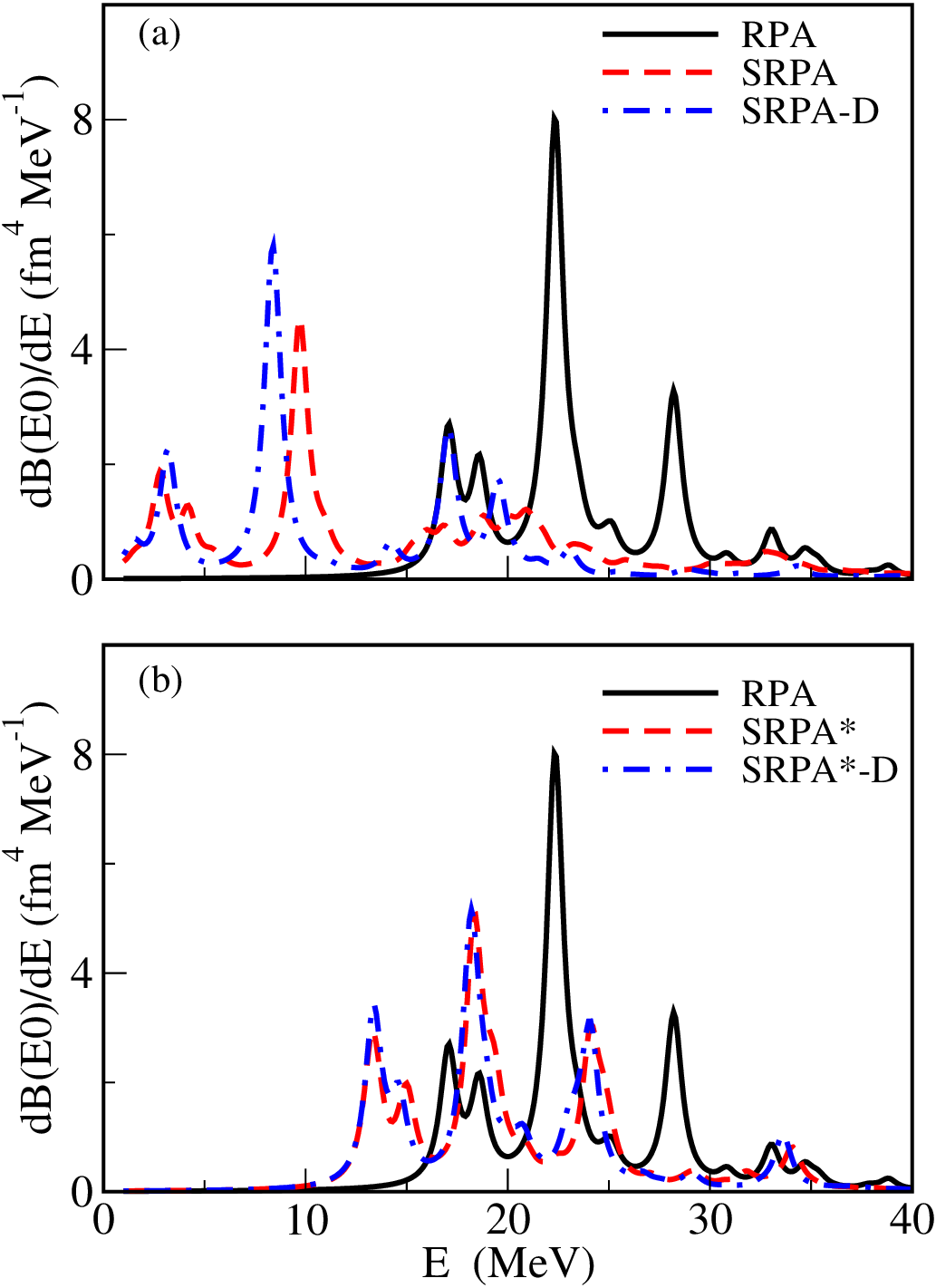} 
\vspace{0.9cm}
\caption{(Color online) Isoscalar monopole distributions for the nucleus $^{16}$O
calculated within the SRPA (a) and SRPA* (b) model (dashed lines) with a
cutoff of 80 MeV. The comparison with the corresponding spectra obtained in the 
diagonal approximation (dot-dashed lines) is shown. 
The Gogny-RPA results are also plotted (full lines).}
\label{fig5}
\end{figure}
%

We have studied the effects of these matrix elements on the transition
densities which are shown in Fig. \ref{fig4} for the energies of the
main peaks of RPA, SRPA and SRPA$^*$ strength distributions. In this
figure, the neutron and proton transition densities are indicated by
the dashed and full lines, respectively.  In the upper panel (a) the
neutron and proton RPA transition densities are plotted for the state
located at $\sim$ 22.34 MeV. In the middle (b) and lower (c) panels we
show the transition densities for the states located at 9.71 and 18.36
MeV, corresponding to the states obtained in the SRPA (b) and SRPA*
(c) cases.  We observe that the shapes of the profiles are rather
similar.  This indicates that the nature of these RPA and SRPA excited
states does not change very much in terms of the spatial distributions
of the contributing wave functions. We conclude that, at least in this
case, the $\nu\pi$ matrix elements affect the energy and the strength
distributions while the shapes of the radial distributions of the
contributing nucleons are not appreciably modified. 

Finally, in the framework of the Gogny-SRPA model we have checked the
validity of the so-called diagonal approximation, which amounts to
neglect the coupling of the 2ph configurations among themselves.  The
results obtained by adopting this approximation will be identified as
SRPA-D. The validity of this approximation has been tested in
Ref. \cite{gamba10} by using the Skyrme interaction. In this case,
large differences with respect to the complete calculations have been
found. In Fig. \ref{fig5} we compare the results obtained in the full
SRPA (a) and SRPA* (b) models with the corresponding results obtained
in the diagonal approximation.  All the calculations have been carried
out by using an energy cutoff of 80 MeV. We observe that in both cases
the shifts and the shapes of the strength distributions do not
strongly change when the residual interaction in the $2ph$ space is
neglected, the differences being larger in the SRPA case than in the
SRPA* one.  The relatively small differences that have been found
between the SRPA and the SRPA-D cases indicate that the effect of
the $\nu\pi$ matrix elements is mainly related to the coupling of the
$1ph$ configurations with the $2ph$ ones.  In the SRPA* case, panel
(b), we see that the results obtained within the diagonal
approximation are extremely close to the full ones and the same
behavior is found also for larger energy cutoff.  By comparing these
results with those obtained by using the Skyrme interaction [panel
  (a) of Fig. 8 of Ref. \cite{gamba10}] we deduce that in the Gogny
case the diagonal approximation provides results which are much closer
to the full results, at least in the monopole case. 

In summary, we have found that in the Gogny-SRPA calculations the
responses are very strongly affected by some $\nu\pi$ matrix elements
of the residual interaction, particularly in the channels which couple
the $1ph$ with the $2ph$ configurations, i.e. 3 hole-1 particle type.
These matrix elements do not
contribute in Hartree-Fock and standard RPA calculations. Therefore,
they do not contribute in the calculations where the parameters of the
effective forces are fixed by the usual fitting procedures.  To check
and constrain their effects, it is thus necessary to go beyond the
conventional procedures.  We suggest some different possible
directions that may be followed to constrain these matrix elements.
First, it is evident that the $\nu\pi$ matrix elements of the residual
interaction play an important role in charge-exchange RPA
calculations. This means that charge-exchange Gogny-RPA calculations
could in principle provide important indications to better constrain
in general this type of matrix elements.  It is also interesting to
notice that the same $\nu\pi$ matrix elements of the residual
interaction entering in the SRPA formalism are also present in the
variational multiparticle-multihole model of Ref. \cite{pillet}.  The
two models could thus be used in a complementary way to analyze the
impact of these terms and to guide toward a better control of their
contributions in beyond mean-field theories.


\end{document}